\documentstyle[eqsecnum,amssymb,prb,aps,multicol,psfig]{revtex}

\begin{document}
\draft 
\title{{\bf A Fluid Dynamics Calculation of Sputtering\\ from a Cylindrical Thermal
Spike}.}
\author{M. M.~Jakas\cite{byline}$^{(a)}$, E. M.~Bringa$^{(b)}$, and R.E. Johnson$%
^{(b)}$}
\address{$^{(a)}$ Departamento de F\'{i}sica Fundamental y Experimental, Universidad\\
de La Laguna, 38201 La Laguna, Tenerife, Spain \\
and,\\
$^{(b)}$Engineering Physics, University of Virginia. Charlottesville, VA\\
22903, U.S.A.}
\date{\today}
\maketitle

\begin{abstract}
The sputtering yield, $Y$, from a cylindrical thermal spike is calculated
using a two dimensional fluid dynamics model which includes the transport of
energy, momentum and mass. The results show that the high pressure built-up
within the spike causes the hot core to perform a rapid expansion both
laterally and upwards. This expansion appears to play a significant role in
the sputtering process. It is responsible for the ejection of mass from the
surface and causes fast cooling of the cascade. The competition between
these effects accounts for the nearly linear dependence of $Y$ with the
deposited energy per unit depth that was observed in recent Molecular
Dynamics simulations. Based on this we describe the conditions for attaining
a linear yield at high excitation densities and give a simple model for this
yield.
\end{abstract}

\begin{multicols}{2}
\narrowtext 

\section{Introduction}

The ejection of atoms from the surface of a solid during ion irradiation is
well documented both experimentally and theoretically \cite{Spikes}. This
phenomenon, known as sputtering, is due to the energy transferred to the
atoms in the target by the incident ion. This produces a cascade which can
cause some atoms to overcome the surface's attractive barrier and escape to
vacuum.

In previous theoretical work the mean number of ejected atoms per incoming
ion, or sputtering yield $Y$, is related to the energy deposited the ion
per unit thickness at the surface of the target $F_{D}$, as $Y\propto
F_{D}^{n}$. The value of the power $n$ depends on the type of collision
cascade produced by the ion, namely linear and non-linear cascades. For
linear cascades, when the density of moving atoms $N_{mov}$ within the
cascade is small compared to normal density $N_{0}$, one has $n=1$ \cite
{Thompson,Sigmund}, whereas in the non-linear case $N_{mov}\sim N_{0}$
theoretical work predicted that $n$ must be greater than one \cite
{Bob-Evatt,S&C}. These results are so firmly established that the consensus
among workers in the field is that $n>1$ and non-linear cascades are to some
extent synonymous \cite{andersen}. Similar results have been found for
sputtering in response to electronic energy deposited in a solid \cite
{johnson-schou}, but here we refer to work on collision cascade sputtering.

Recent Molecular Dynamics (MD) studies \cite{Urbassek-Bob,Edu-Bob}, however,
cast doubt on this relationship. According to these papers, purposely
prepared non-linear cascades can give rise to sputtering yields which depend
linearly on $F_{D}$\ (see figures \ref{Y-for-lambda}-\ref{Y-for-c}). Further
evidence is found in Ref.\cite{PRB1}. After modifying the standard thermal
spike theory (STST) to include the transport of mass, the sputtering yields
so calculated appeared to be much closer to a linear function of $F_{D}$
than to the $F_{D}^{2}$ predicted by the STST.

Although the results in Ref.\cite{PRB1} show the importance of having a
target which can change its specific volume as a fluid, it is not a full
fluid dynamics calculation. Since the target was assumed to be infinite,
the sputtering yields had to be calculated in the same manner as in the
STST. That is, an expression for the evaporation rate was used that was
borrowed from the kinetic theory, and the sputtering yields were obtained by
integrating it along a plane representing an otherwise non-existent surface.
Further, the transport was only radial, but the MD calculations showed the
importance of energy transport along the track.

In order to circumvent such a difficulty, our previous calculations are
extended to a target which, in addition to being compressible, has a
solid-vacuum interface. To this end, the target density, velocity and
internal energy are all assumed to vary with time in a manner which is
described by the fluid dynamics equations. Consequently, sputtering emerges
naturally, as that part of the target that succeeds in escaping from the
condensed to the gaseous phase.

The aim of this paper is to show the most relevant aspects of those
calculations, from the underlying mathematics to the results and
implications of the proposed model. Although the present calculations can be
applied to a variety of ion-induced thermal spike cases, we have purposely
limited ourselves to the cases contained in previous MD simulations \cite
{Urbassek-Bob,Edu-Bob}. Therefore, the results in this paper are intended
for cylindrical thermal spikes. In Section \ref{Theory} we introduce the
fluid dynamics equations as well as the various expressions used along the
present calculations. Results and discussions are presented in Section \ref
{Results and discussions}. Finally, the conclusions and suggestions for
further studies are presented in Section \ref{Conclusions}.

\section{Theory}

\label{Theory}

We assume that the target is a continuous medium with cylindrical symmetry,
and it is completely characterized by its atomic number density $N$,
velocity ${\bf v}$, and internal energy $\epsilon $ (per atom) defined as,

\begin{equation}  \label{Epsilon}
\epsilon =U+\frac{3}{2}k_{B}T,
\end{equation}

\noindent where $k_{B}$ is the Boltzmann's coefficient, $T$ the temperature
and $U$ is the potential energy per atom. It must be noted that by using the
equation above the heat capacity at constant volume, $C_{V\text{ }}$, is
assumed to be that of a dilute gas, i.e. $3k_{B}/2$. This approximation
however is fairly acceptable for the purpose in this paper, since as shown
in Ref.\cite{MJakas}, the quadratic dependence of $Y$ with $F_{D}$ does not
appear to be connected to $C_{V\text{ }}$. Moreover $U$ is obtained from the
expression \cite{PRB1}

\begin{equation}  \label{UBIND}
U = (N_{0}M c_{o}^{2}/\mu )\ (N/N_{0})^{\nu+\mu-1 }\ \left[\frac{1}{\nu+\mu-1%
}-\frac{(N/N_{0})^{2}}{\nu+\mu+1}\right] \ ,
\end{equation}

\noindent where $M$ is the mass of the target atom, $c_{o}$ is the speed of
sound at $T=0K$, and $N_{0}$ is the normal atomic number density. $\mu $\
and $\nu $\ are two numerical constants which, as we explained in Ref.\cite
{PRB1}, are not independent. Thus we set $\mu =2$, then $\nu =\sqrt{%
1+Mc_{0}^{2}/U_{0}}$, $U_{0}$ being the potential energy at normal density,
i.e. $U_{0}=-U(N_{0})$.

Using the same notation as in Ref. \cite{Landau}, we write the fluid
dynamics equations as follows

\begin{equation}  \label{Cont.}
\frac{\partial N}{\partial t} = - \frac{\partial (v_{k}N)}{\partial x_{k}}\ ,
\end{equation}

\begin{equation}  \label{moment}
\frac{\partial v_{i}}{\partial t} = - v_{k} \frac{\partial v_{i}}{\partial
x_{k}} - \frac{1}{NM} \left( \frac{ \partial p}{\partial x_{i}} + \frac{%
\partial \sigma^{\prime}_{ik}}{\partial x_{k}} \right)
\end{equation}

\begin{equation}  \label{energy}
\frac{\partial \epsilon}{\partial t}= - v_{k}\frac{\partial \epsilon}{%
\partial x_{k}} + \frac{1}{N} \left( Q_{con} + Q_{vis} - p \frac{\partial
v_{k}}{\partial x_{k}} \right) \ ,
\end{equation}

\noindent where the subscripts stand for the $r$ and $z$ coordinates, $p$ is
the pressure and $\sigma^{\prime}_{ik}$ is the viscosity tensor \cite{Landau}
defined as,

\begin{equation}  \label{Visc.tensor}
\sigma^{\prime}_{ik} = \eta \left(\frac{\partial v_{i}}{\partial x_{k}} + 
\frac{\partial v_{k}}{\partial x_{i}}\right)\ ,
\end{equation}

\noindent where $\eta $ is the {\em dynamic viscosity} coefficient and $%
Q_{con}$ and $Q_{vis}$ account for the heat produced by thermal conduction
and viscosity per unit volume and time, namely

\begin{equation}  \label{ThermCond}
Q_{con} = {\bf \nabla} \left(\kappa_{T}{\bf \nabla}T\right)\ ,
\end{equation}

\noindent where $\kappa_{T}$ is the thermal conductivity and,

\begin{equation}  \label{VisHeat}
Q_{vis} = \sigma^{\prime}_{ik} \frac{\partial v_{i}}{\partial x_{k}} \ .
\end{equation}

The heat conduction coefficient is replaced by that in Ref. \cite{S&C}

\begin{equation}  \label{Eq.3}
\kappa_{T}=\frac{25}{32}\frac{k_{B}}{\pi a^{2} }\sqrt{\frac{k_{B}T}{\pi M}}\
,
\end{equation}

\noindent where $\pi a^{2}=1.151$\AA $^{2}$. This form was used in order to
compare results and because there seems to be no reason for using a more
``realistic'' one since, as shown in Ref. \cite{MJakas}, $\kappa _{T}$ and
the quadratic dependence of the sputtering yield appear not to be connected.

Making use of the fact that, for dilute gases, $\eta $ and $\kappa _{T}$ are
related through the equation $\eta =\kappa _{T}M/(3k_{B})$, we thus
introduce the dimensionless viscosity coefficient

\begin{equation}  \label{eta}
\eta^{*} = 3 k_{B} \eta/(M \kappa_{T})\ .
\end{equation}

Similarly, the pressure $p$ is assumed to be a function of both temperature
and density. Here, we follow the approximation in Ref. \cite{Zeldovich} and
split $p$ into two terms

\begin{equation}  \label{Ptotal}
p = p_{T} + p_{C}\ ,
\end{equation}

\noindent where the {\em thermal pressure} $p_{T}$ can be obtained from the
expression

\begin{equation}  \label{Ptherm}
p_{T}= \lambda\ Nk_{B}T,
\end{equation}

\noindent $\lambda $ being a numerical constant. The so called {\em crystal
pressure} $p_{C}$ can be obtained from the potential energy Eq.(\ref{UBIND})
using the equation \cite{Zeldovich}

\begin{equation}  \label{Pxtal}
p_{C} = N^{2}\frac{\partial U}{\partial N}\ .
\end{equation}

\begin{figure}[tbp]
\centerline{\psfig{figure=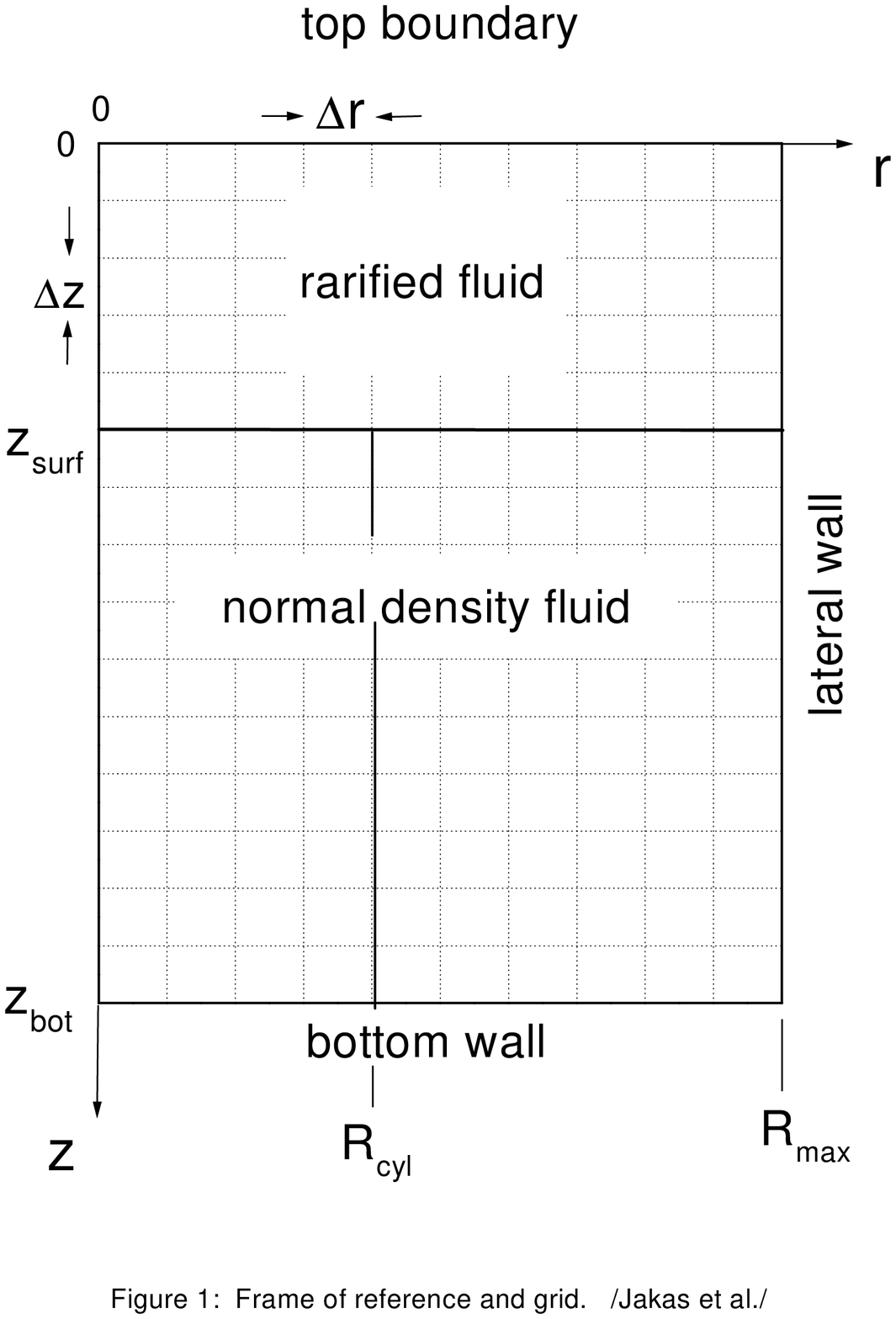,width=6.5cm,height=8.5cm,angle=0}}
\caption{ Sketch illustrating the frame of reference and grid
utilized in the present calculations. At $t$=0 the ``fluid'' occupies 
the region defined by inequalities $z_{surf}<z<z_{bot}$ and $0<R<R_{max}$, 
and the hot spike is confined to a cylinder of radius $R_{cyl}$.}
\label{System}
\end{figure}

For computational purposes, Eqs. (\ref{Cont.}-\ref{energy}) are applied to a
finite system, which is defined by inequalities $0\leq r\leq R_{max}$ and $%
0\leq z\leq z_{bot}$ (see figure \ref{System}). Furthermore, the top wall,
i.e. $z=0$, is assumed to be made of a perfectly absorbent material, whereas
the boundary at the bottom is perfectly closed as far as to the exchange of
mass, momentum and energy is concerned. The lateral wall can be made either
closed, like the bottom surface, or partially open. That is, closed to mass
transport but open to energy and momentum exchange. Results in this paper
were obtained using the latter option. Otherwise, in the case of a large
deposited energy, one would need an exceedingly large target to minimize the
effects of energy and momentum reflections. A more detailed description of
this program will be published elsewhere \cite{HYDRO2D}.

At $t=0$ the target is within a range of $z$ defined by inequality $(z\geq
z_{surf})$. For numerical reasons however, we assume that the region that
would normally be a vacuum is filled with a low-density fluid, i.e. $%
N_{min}=10^{-3}\times N_{0}$. Exchange of energy, momentum and matter is 
forbidden in this fluid, as well as in any other piece of a fluid with 
density lower than $N_{\min }.$ The possible net flux of matter is 
continuously checked along the fluid, and the restrictions above are relaxed 
as soon as an element of the fluid would have its density increased above 
$N_{min}$.

To energize the spike, all atoms within a cylinder of radius $R_{cyl}$ are
given an energy $E_{exc}$ above their initial energy, $\epsilon
_{0}=-U_{0}+(3/2)k_{B}T_{0}$, where $T_{0}$ is the background temperature,
often assumed to be 10 K. These are the initial conditions used in a number
of the MD simulations \cite{Urbassek-Bob,Edu-Bob}, again allowing direct
comparison with the results here. The initial conditions for Eqs.(\ref{Cont.}%
-\ref{energy}) thus become,

\begin{eqnarray}  \label{init.conds.}
v_{r,z}(0,r,z) &=&0\ ,  \nonumber \\
N(0,r,z) &=&\left\{ 
\begin{array}{l}
N_{0}\mbox{~~~if~~~ }z\geq z_{surf} \\ 
N_{min}\mbox{~~~~otherwise}
\end{array}
\right. \\
\epsilon (0,r,z) &=&\left\{ 
\begin{array}{l}
E_{exc}+\epsilon _{0}\mbox{~~~~~if~ }r\leq R_{cyl}\mbox{ and }z\geq z_{surf}
\\ 
U(N_{min})+(3/2)k_{B}T_{0}\mbox{~~~~~if~}0\leq z<z_{surf} \\ 
\epsilon _{0}\mbox{~~~~otherwise}
\end{array}
\right.  \nonumber
\end{eqnarray}

With the assumptions above, the deposited energy becomes,

\[ F_{D}=\pi R^{2}_{cyl}N_{0}E_{exc}\ . \]

As is customary, in solving the fluid dynamics equations the functions $N$, $%
{\bf v}$ and $\epsilon $ are defined over a discrete set of $NR\times NZ$
points, whose mesh size is determined by $\Delta r$ and $\Delta z$ (See
figure \ref{System} and Table \ref{Table_I}). A compromise has to be made
about target size since a large target implies a fairly large
system of coupled equations with fairly long running times. Whereas too
small a target gives rise to boundary effects that would make calculations
meaningless. Similarly, in choosing $z_{surf}$ one has to take into account
that during ejection not all the matter that crosses the surface will be
ejected. Therefore, the distance between the initial surface and the top
wall should be large enough to not ``collect'' matter that, otherwise, would
not be ejected. Finally, the piece of matter representing the target must be
thick enough. The condensed phase is assumed to be 10$\sigma$ thick, which
means that $z_{surf}\approx$10$\sigma$\ and $z_{bot}\approx$20$\sigma $. $%
NR=40$\ and $NZ=20$\ were found to be adequate for all the cases studied in
this paper.

When integrating the fluid dynamics equations (\ref{Cont.},\ref{moment},\ref
{energy}) from $t=$0 to $t_{f}$, the total flux of matter passing through
the top boundary is also calculated. In this way the sputtering yield is
obtained as a function of time, $Y(t)$. This is used to verify if $t_{f}$
was long enough so that no matter remains within the system that may
significantly contribute to the sputtering yield. We use the $Y(t)$'s for $%
t<t_{f}$\ to extrapolate $Y(t)$ to infinity, i.e. $Y_{\infty
}=\lim_{t_{f}\rightarrow \infty }Y(t_{f})$. Only runs for which $Y_{\infty
}-Y(t_{f})\approx 0.1Y_{\infty }$ are accepted. Normally, $t_{f}$ ranging
from 10 up to 50 ps are required.

Since calculations in this paper are meant to be compared with those in MD
simulations, which often use Lennard-Jones (LJ) potentials, the various
parameters characterizing our system correspond to those of Argon 
(see Table \ref{Table_I}). $M$=40u and $U_{0}$=0.08eV have become standard 
parameters \cite{Urbassek-Bob,Edu-Bob} although the LJ calculations fully 
scale with $U_{0}$ and $M$. Therefore, the results apply to a broader set 
of materials as shown also using a Morse potential\cite{BJJ}. Consistent 
with this, for most cases we used $\Delta r=\Delta z=\sigma$, where 
$\sigma$\ is the LJ distance. However, as several approximations were 
introduced, we cannot ensure that the fluid in our calculations accurately 
describes the potentials used in the MD simulations. Similarly, we do not 
want to leave this section without mentioning that although the fluid 
representing the target is assumed to be compressible, Eq.(\ref{Visc.tensor}) 
looks the same as that of an incompressible fluid because the Stokes' 
condition is assumed to hold.
\begin{table}[tbhp]
\caption{Value of the parameters used in the present calculations.}
\label{Table_I}
\begin{tabular}{c|c|c}
Property & Symbol & Value \\ \hline
Atomic mass & $M$ & 40.0 u.m.a. \\ 
Atomic number density & $N_{0}$ & 0.026 at/\AA$^{3}$ \\ 
Speed of sound & $c_{0}$ & 17 \AA/ps \\ 
Binding energy & $U_{0}$ & 0.08 eV \\ 
Lennard-Jones distance & $\sigma$ & 3.405 \AA
\end{tabular}
\end{table}

\section{Results and discussions}

\label{Results and discussions}

\begin{figure}[tbp]
\centerline{\psfig{figure=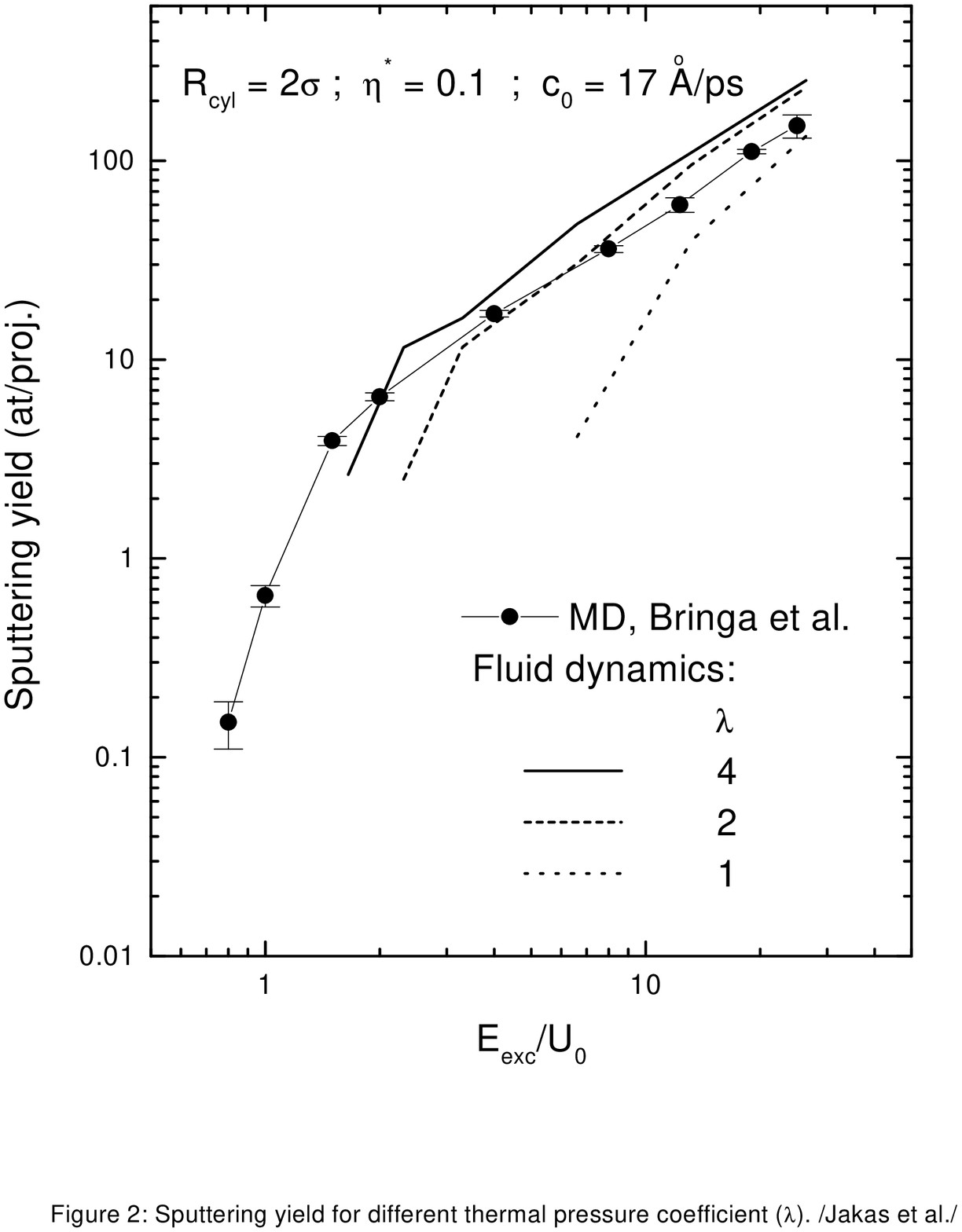,width=7.cm,height=9.5cm,angle=0}}
\caption{ Sputtering yield as a function of the excitation energy and 
different values of parameter $\lambda $.}
\label{Y-for-lambda}
\end{figure}

\begin{figure}[tbp]
\centerline{\psfig{figure=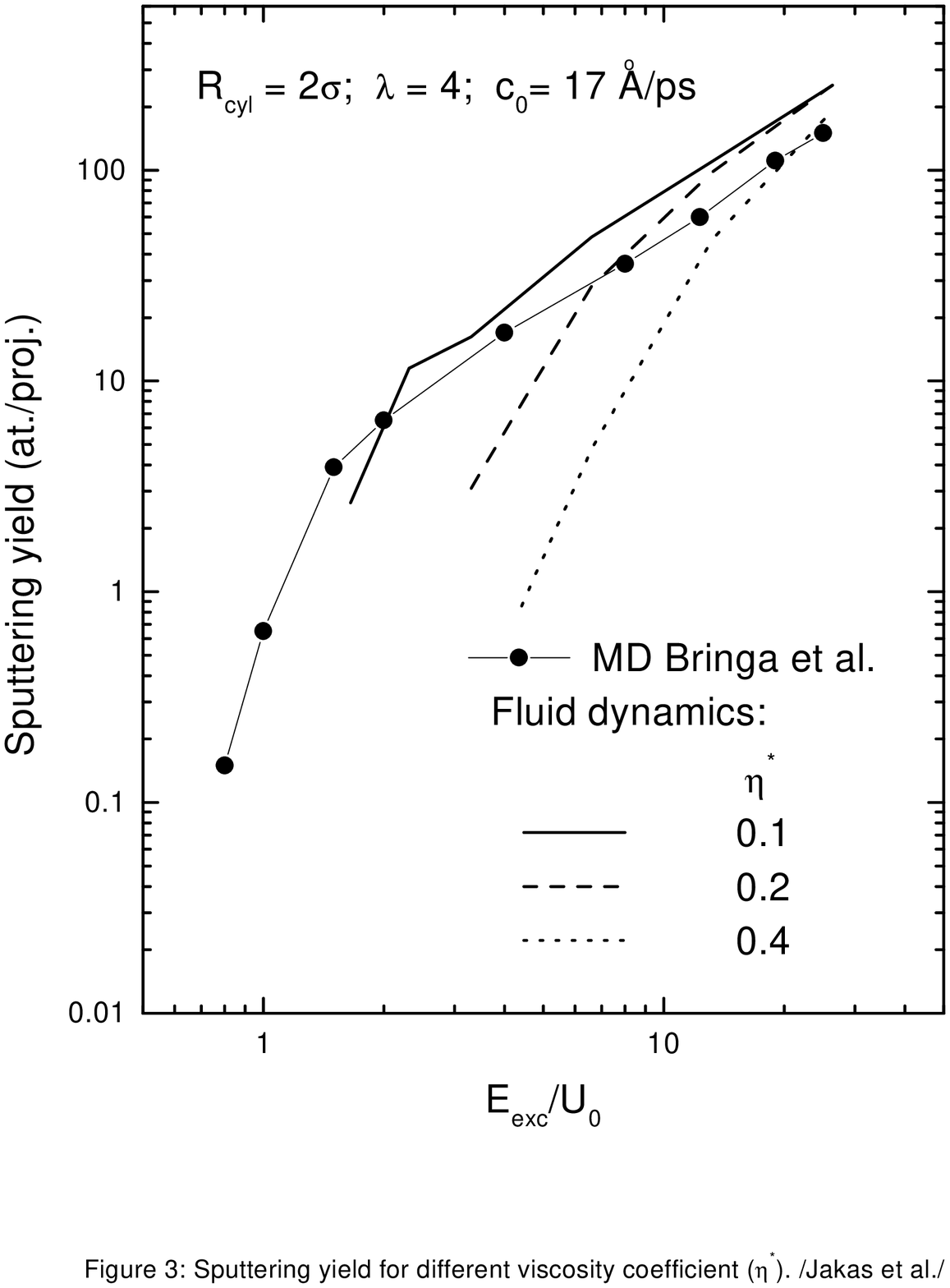,width=7.cm,height=9.5cm,angle=0}}
\caption{ Sputtering yield as a function of the excitation
energy and different values of viscosity coefficient $\eta ^{*}$.}
\label{Y-for-eta}
\end{figure}

\begin{figure}[tbp]
\centerline{\psfig{figure=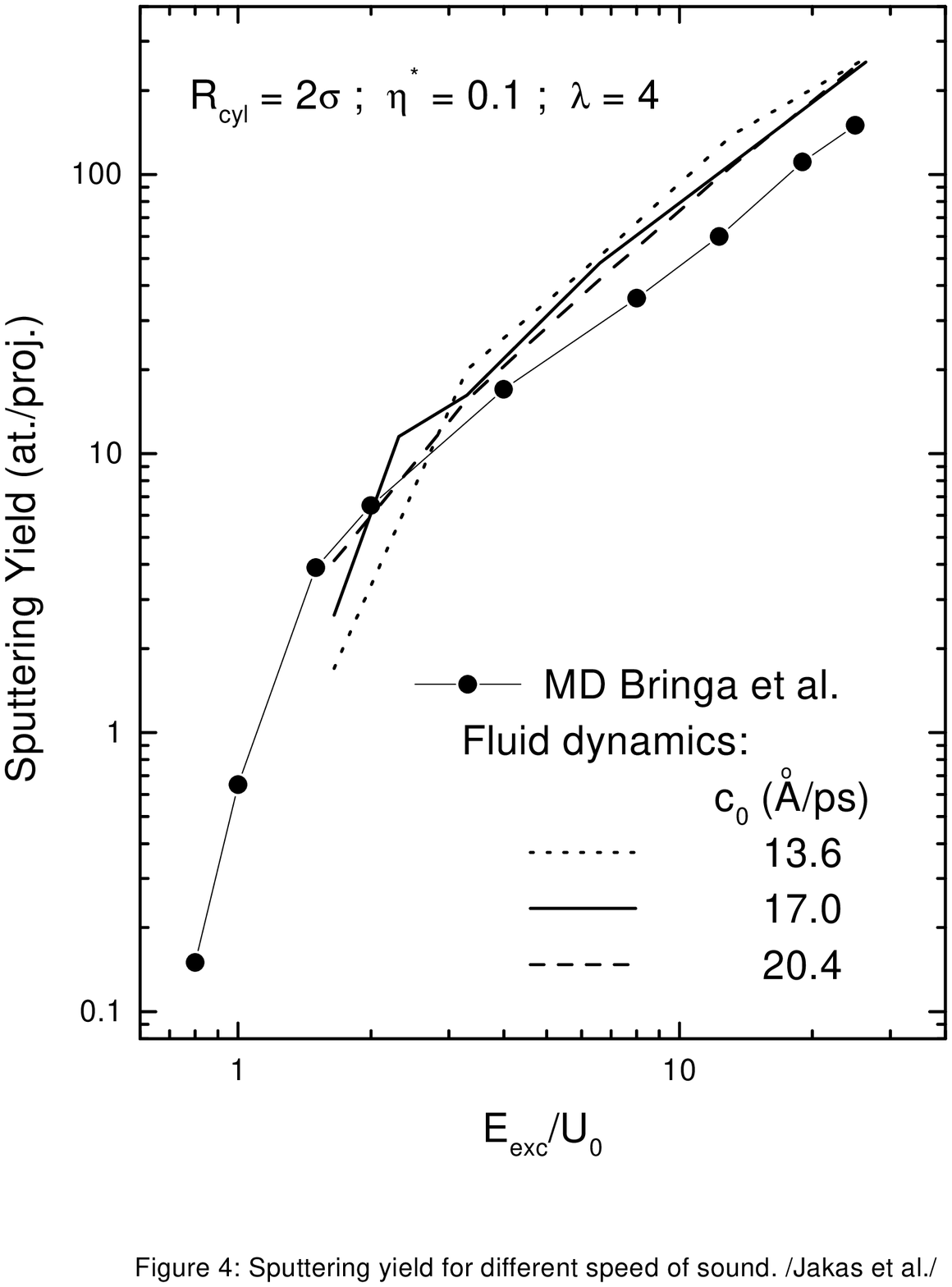,width=7.cm,height=9.5cm,angle=0}}
\caption{ Sputtering yield as a function of the excitation energy and 
different values of the speed of sound $c_{0}$.}
\label{Y-for-c}
\end{figure}

We calculated the sputtering yield for different values of $\lambda $, 
$\eta^{*}$ and the speed of sound $c_{o}$, and the results are depicted in
figures \ref{Y-for-lambda} to \ref{Y-for-c}. We observe that in all the
cases the yield increases with increasing excitation energy $E_{exc}$.
Similarly, $E_{exc}\approx U_{0}$ is an effective threshold for ejection
for the initial radius used, since the yields rapidly decrease for $E_{exc}$
comparable to or less than $U_{0}$. Whereas the MD requires varying
potential types to obtain different material properties, here we do this by
directly varying the material properties. In this manner the relationship
between different materials can be described.

We observe that $\lambda $ has a great influence on the sputtering yield.
The larger the $\lambda $ the greater the yield. $\lambda $=4 appears to
reproduce MD-simulations quite well, whereas $\lambda $=2 and 1
underestimate the yields at small excitation energies. These results are, to
some extent, easy to understand: with all other parameters remaining the
same, as $\lambda $ becomes larger the thermal pressure build up within the
spike increases and more ejection is expected.

It must be noted, however, that the total energy, c.f. Eq.(\ref{Epsilon}),
does not depend on $\lambda $. Increasing $\lambda $\ only increases the
thermal pressure and speeds up the conversion of thermal motion into
directed kinetic energy. Therefore, thermal conductivity has less time to
take energy away from the spike and the ejection of matter increases.

The role played by the viscosity on the sputtering yield is illustrated in
figure \ref{Y-for-eta}. For the cases studied here, we observed that
viscosity has a negative influence on the ejection process, as yields are
seen to get smaller with an increase of the viscosity coefficient. At small
excitation energies the viscosity appears to play a major role. Furthermore,
calculations using $\eta ^{*}=0.1$ produced a good agreement with
MD-simulations while those with $\eta ^{*}=0.2$ and $0.4$ resulted in
significantly smaller yields. The fact that the best agreement with
MD-simulations corresponds to calculations with $\eta ^{*}=0.1$ is not
unexpected since $\eta ^{*}$-values of approximately that order have been
calculated for a Lennard-Jones fluid \cite{Hisrchfelder}.

Finally, modifying the speed of sound does not produce a significant change
in the sputtering yield. Figure \ref{Y-for-c} shows results for the speed of
sound both above and below its normal value. The change in the sputtering
yield is very small compared that produced by changing either the viscosity
coefficient or the thermal pressure coefficient $\lambda $. We observe that,
for high excitation energies, an increase in the speed of sound leads to a
slightly greater yield, and that such a trend is reversed as $E_{exc}/U_{0}$
becomes smaller than 3.

\begin{figure}[tbp]
\centerline{\psfig{figure=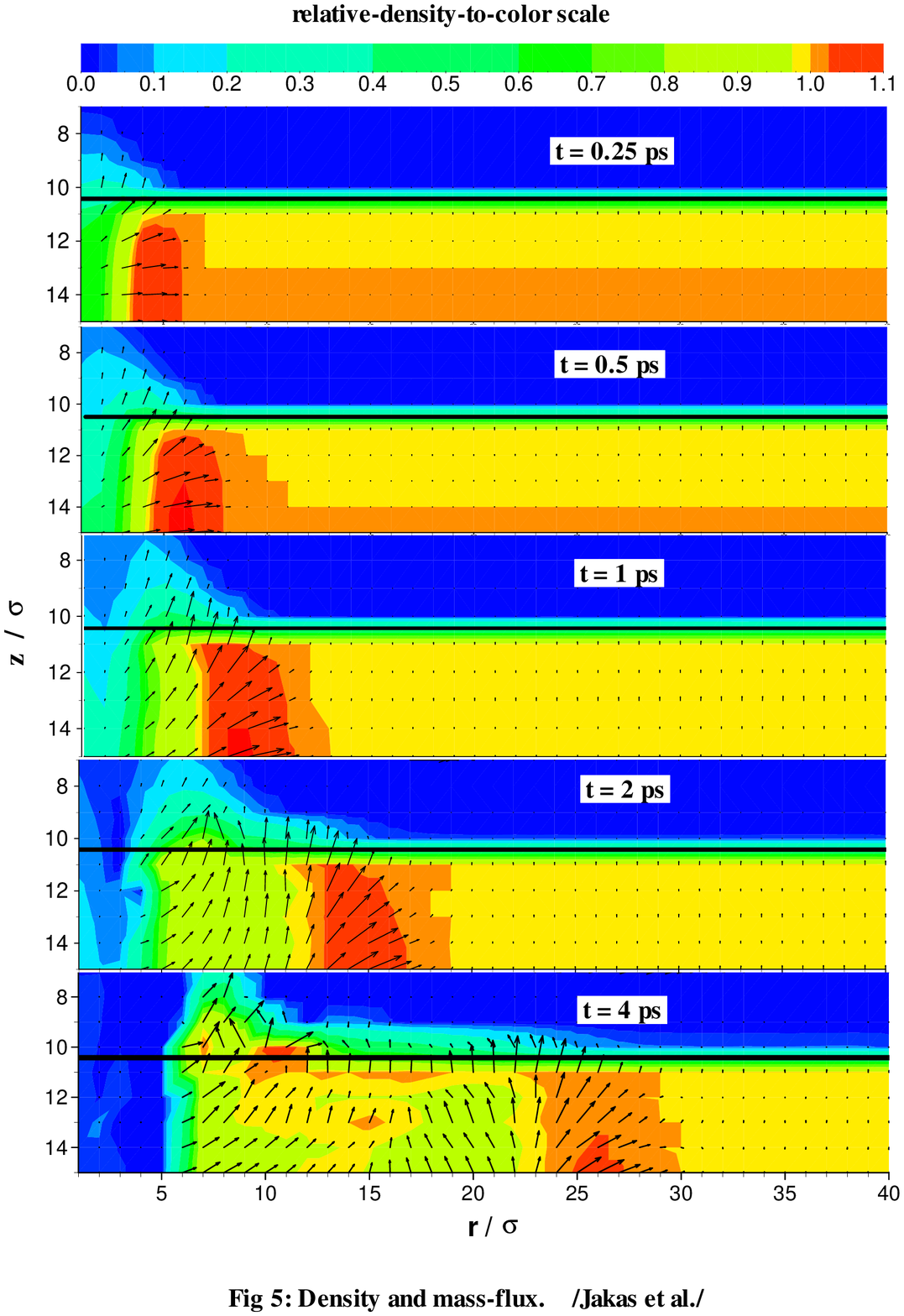,width=7.0cm,height=10.cm,angle=0}}
\caption{ These plots illustrate the density and
mass-flux vectors at different times for a spike with $dE/dX$=4eV/\AA , $%
\lambda $=4, $\eta ^{*}$=0.1, $c_{0}$=17 \AA /ps and $R_{cyl}$ = 2$\sigma $.
The scale used for translating from relative density ($N/N_{0}$) into colors
is shown up in the figure. Note that the scale is non-linear, as more colors
are used at both small densities and around $N/N_{0}$=1. Furthermore, due to
interpolation in the plotting software, details of the order of the grid
size, or smaller, might not be accurately copied. The horizontal line
denotes the initial position of the surface.}
\label{N-and-V-of-t}
\end{figure}

\begin{figure}[tbp]
\centerline{\psfig{figure=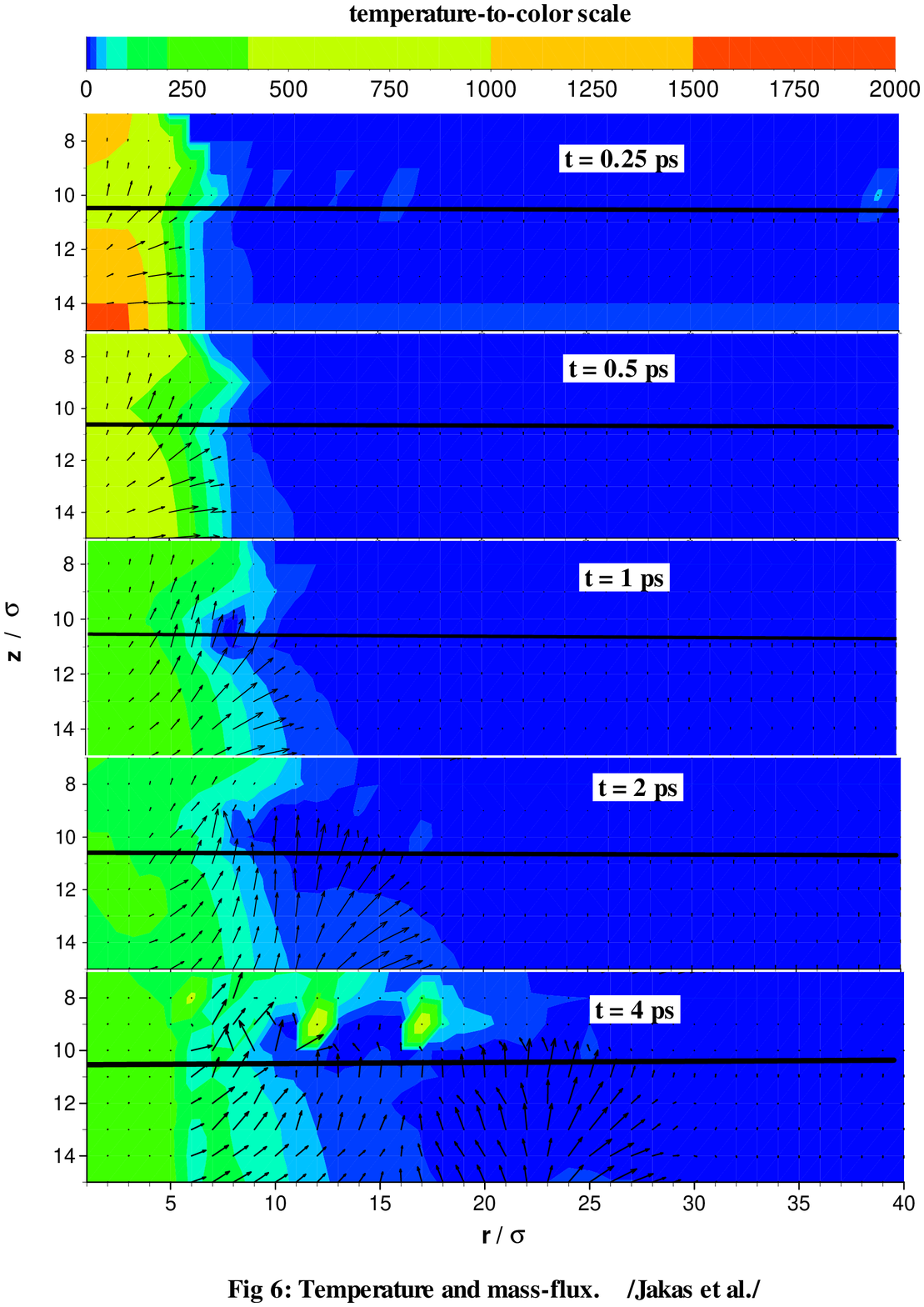,width=7.cm,height=10.cm,angle=0}}
\caption{ Temperature and mass-flux vectors at different
times within the fluid for the spike described in caption 
\protect{\ref{N-and-V-of-t}}. The scale used for color vs. temperature 
appears up in the figure.}
\label{T-and-V-of-t}
\end{figure}

\begin{figure}[tbp]
\centerline{\psfig{figure=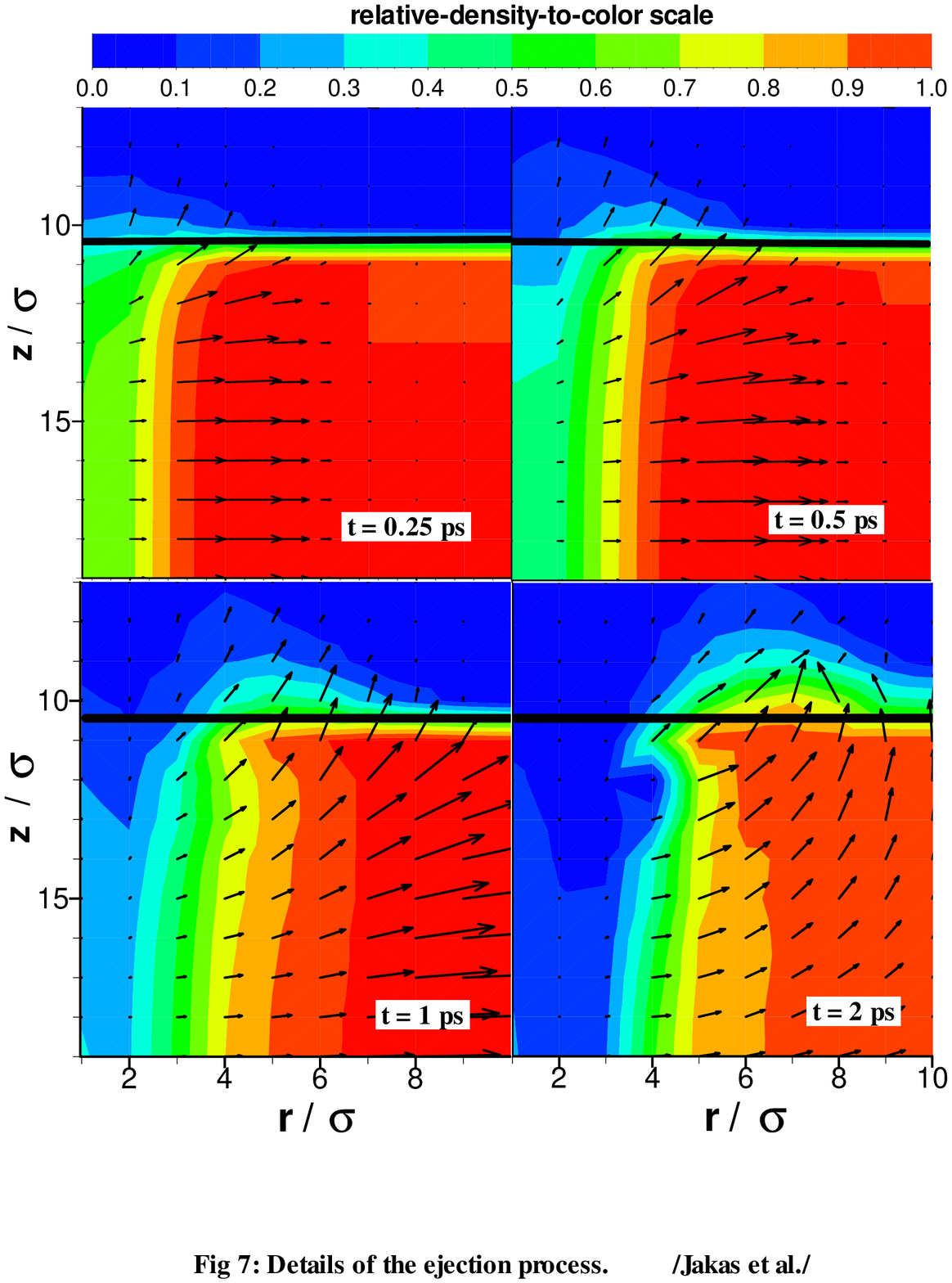,width=7.cm,height=10.cm,angle=0}}
\caption{ Close-ups of plots in figure 
\protect{\ref{N-and-V-of-t}} illustrating in more details the dynamics 
of the fluid within the ``core'' of the spike and near the surface.}
\label{N-and-V-of-t(2)}
\end{figure}

As we mentioned in the introduction, the most interesting result in this
paper is our ability to explore the material parameters that lead to the
near linearity exhibited in the yield in our MD calculations even though the
sputtering is a non-linear process. By exploring the parameter space we can
better explain that phenomenon and assess its relevance. Our calculated
yields in figures \ref{Y-for-lambda}-\ref{Y-for-c} clearly show that a
linear region is attained for $E_{exc}>U_{0}$ using a set of materials
parameters. Therefore, non-linear sputtering does not necessarily imply
non-linear yields. From these figures, it also appears that linearity is
approached at higher energy densities than those studied here for other
materials parameters. Below we describe this phenomenon.

To understand the change in the dependence of the yield with increasing
excitation density for fixed $R_{cyl}$, we analyzed the time-evolution of
the spike paying particular attention to those aspects of the energy and
momentum transport that are related to the ejection of matter. To this end,
in figures \ref{N-and-V-of-t}-\ref{T-and-V-of-t} we have plotted the
density, the mass-flux vector and temperature in the fluid at different
times after the on-set of the spike. These cases correspond to a deposited
energy of 4 eV/\AA, $\lambda =4$, $\eta ^{*}$=0.1, $c_{0}$=17 \AA /ps
and $R_{cyl}$=2$\sigma $; and, in the three figures, the initial surface is
located at $10\sigma $, i.e. $z_{surf}=10\sigma $.

One readily observes that the temperature within the spike drops below 500 K
in approximately 1 ps, and that the fluid immediately surrounding the spike
hardly reaches temperatures higher than, say, 100 K. This is in agreement
with our MD results and our earlier fluid dynamics calculations \cite{PRB1}.
These studies already showed that, due to the quick, adiabatic expansion of
the fluid, the temperature of the spike decreases much more rapidly than it
would due to thermal conduction. In addition, for times greater than 1 ps
the thermal energy appears to be converted into an elastic wave (seen as red
in Fig. \ref{N-and-V-of-t}) that travels in the radial direction at
approximately the speed of sound. The reader must be aware of the non-linear
scale used in Fig. \ref{N-and-V-of-t} where colors were purposely chosen so
as to change rapidly around both $N_{0}$ and at low density. Due to this,
even the rather small relaxation of the surface density appears as a yellow
stripe, which extends to the right of the spike and gets thicker with
increasing time. These figures show us that the whole process would be
better described as an ``explosion'' rather than a smooth, thermally
diffusive release of energy as proposed in the STST \cite{S&C}.

Note that, in contrast to material further away from the surface, the fluid
that is near the surface and within the spike, appears to follow a
spherical, rather than a cylindrical expansion. That is, if one interpolates
the mass-flux vector and figures out the streamlines of the fluid, then, one
can readily see that near the open boundary of the spike, they seem to
radiate out from a point located on the spike axis somewhere below the
surface. In order to understand this, one has to realize that the momentum
acquired by any particle within the fluid results from the fast, though
small, displacements of the lateral and top boundaries which takes place at
an earlier stage of the aforementioned explosion.

The forces produced by such displacements propagate at the speed of sound
which, within the hot spike, is faster than $c_{0}$\cite{c}. Therefore, by
the time all the fluid within the spike has been set into motion, i.e. $%
t=R_{cyl}/c$ after the onset of the spike, a particle at $(r,z)$ with 0$\leq
z\leq R_{cyl}$ and 0$\leq r\leq R_{cyl}$ will have acquired a velocity that
is proportional to the time it has been exposed to such forces, namely $%
v_{r}\propto r$ and $v_{z}\propto -(R_{cyl}-z)$. Therefore, as $%
v_{z}/v_{r}\approx -(R_{cyl}-z)/r$ this particle will appear as moving away
from a point located exactly on the axis at R$_{cyl}$ below the surface. By
the same token, any particle at a depth greater than $R_{cyl}$ within the
spike, will remain unaware of the presence of the surface and its velocity
will be directed along the radial direction (see figure \ref{N-and-V-of-t(2)}%
). With increasing time our description above will become less and less
accurate. However, as the forces acting within the spike take the largest
values during the earliest stage of the ``explosion'', the velocities
achieved by the fluid during that time essentially determine the
subsequent dynamics of the spike.

Another aspect of the velocity field which deserves attention is that around
the rim, on the cold side of the spike. Contrary to what happens deep in the
target, where the cold side is compressed and subsequently displaced along
the radial direction, the rim is partially wiped out. This not only adds
more matter to sputtering per se, but also clears the way for further
ejection as it widens the sputtering radius, or the radius from within which
particles are ejected.

\begin{figure}[tbp]
\centerline{\psfig{figure=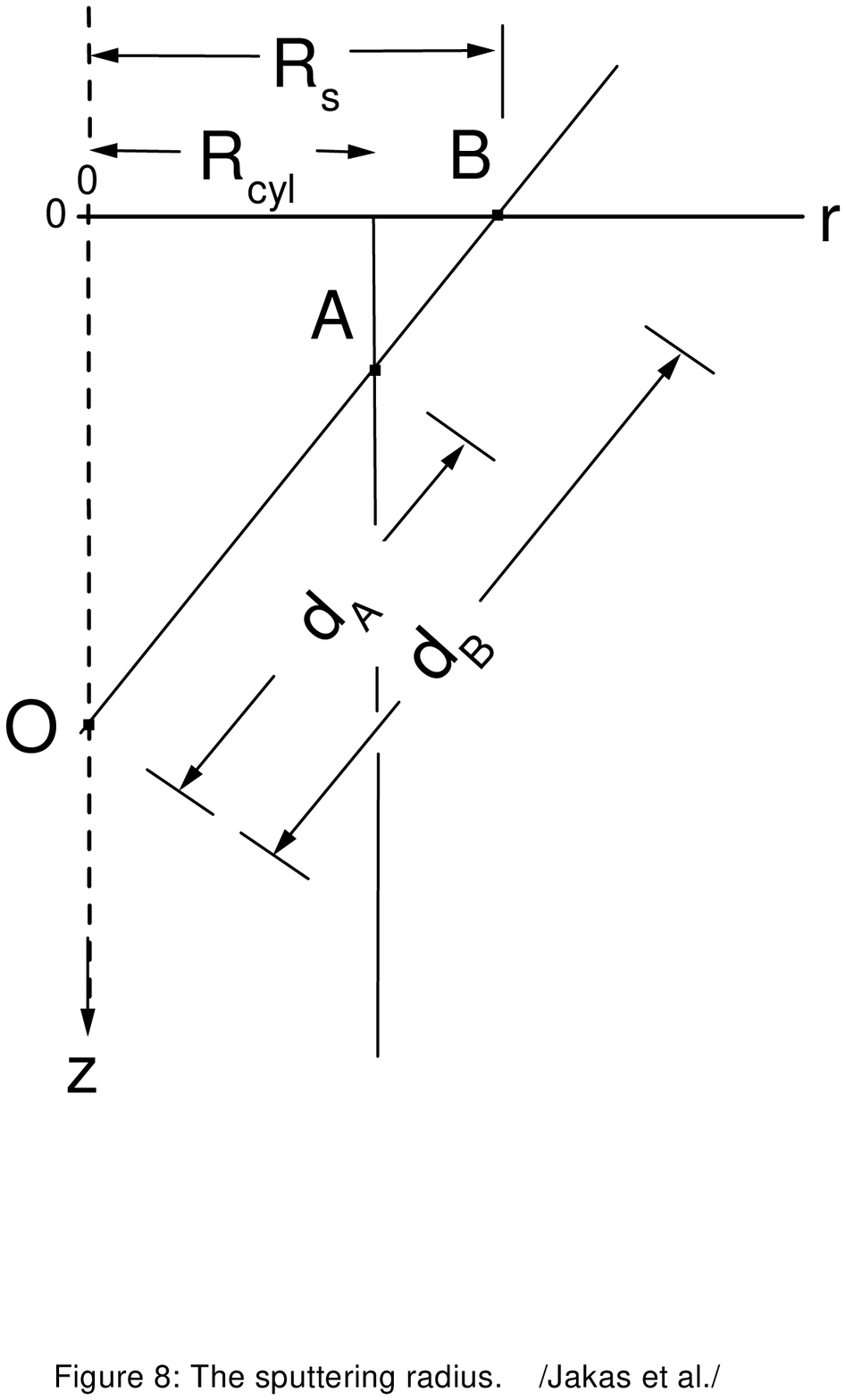,width=7.cm,height=9.cm,angle=0}}
\caption{ Schematics used to obtain the sputtering radius.}
\label{SimMod}
\end{figure}

From this simple picture one can readily calculate the sputtering radius. To
this end, we define the excess energy as the total energy per particle
relative to the bottom of the potential well, i.e. $e=\epsilon +\frac{1}{2}%
Mv^{2}+U_{0}$. If one assumes that the elastic wave in the upper part of the
spike propagates isentropically along the streamlines, one may write

\begin{equation}  \label{App1}
e_{A}/d^{2}_{A} = e_{B}/d^{2}_{B}\ ,
\end{equation}

\noindent where $d_{A}$ and $d_{B}$ are the distance from the center of the
spherical expansion to points A and B, respectively (see figure \ref{SimMod}%
); similarly, $e_{A}$ and $e_{B}$\ are the corresponding excess energies.
Therefore, as $e$ $\geq $U$_{0}$ is a necessary condition for ejection, $%
e_{A}=E_{exc}$ and $R_{cyl}/d_{A}=R_{B}/d_{B}$, one can obtain the
sputtering radius ($R_{s}$) as\cite{U&S} 

\begin{equation}
R_{s}\approx R_{cyl}\left( E_{exc}/U_{0}\right) ^{1/2}\ ,  \label{App2}
\end{equation}

Accordingly, the sputtering yield can be calculated as the amount of mass
contained within a cone of height $R_{cyl}$ and base radius $R_{s}$, i.e.

\begin{equation}
Y\approx \frac{\pi }{3}NR_{cyl}^{3}\frac{E_{exc}}{U_{0}}\ ,  \label{App3}
\end{equation}

\begin{figure}[tbp]
\centerline{\psfig{figure=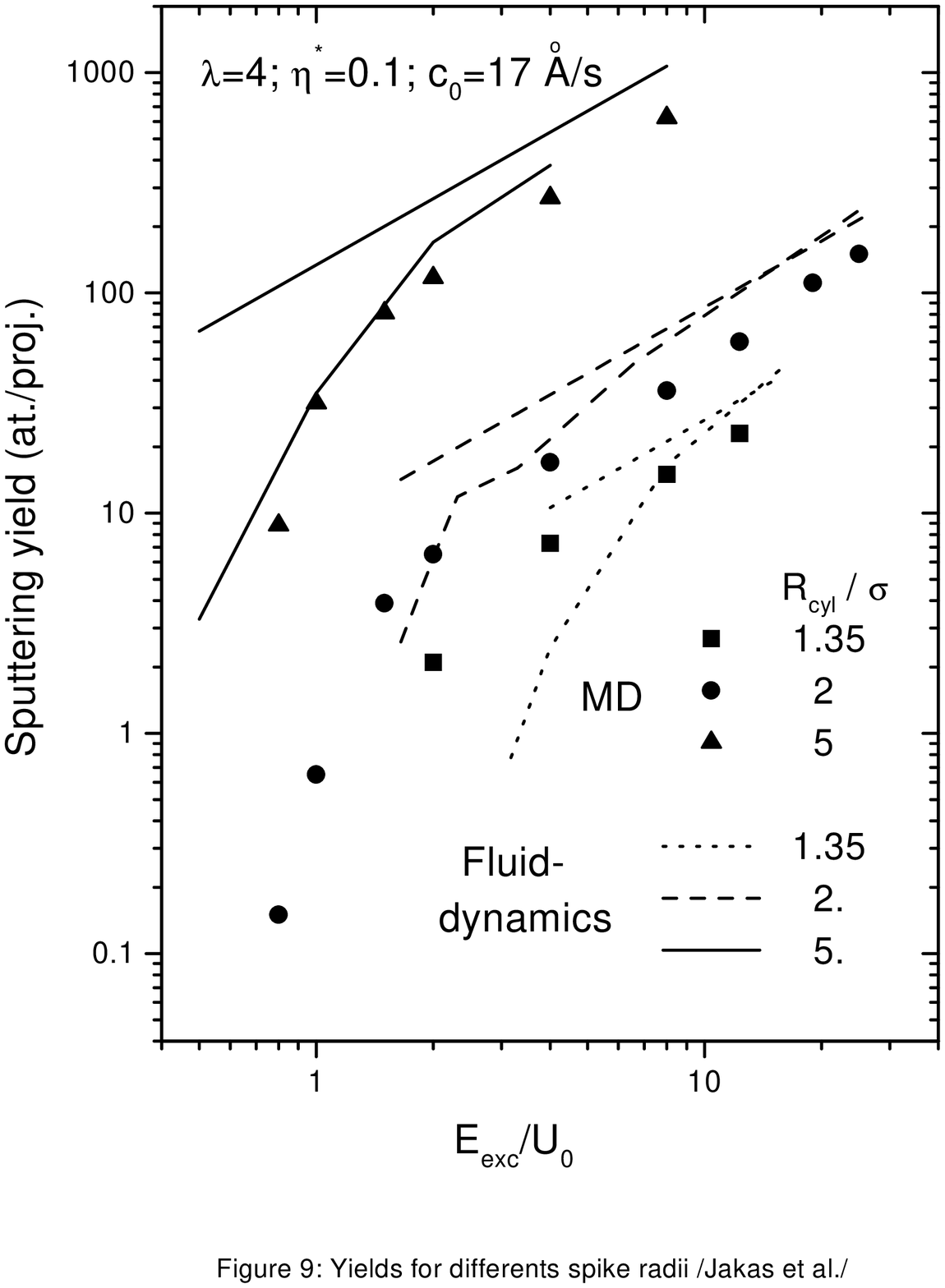,width=7.5cm,height=9.5cm,angle=0}}
\caption{ Sputtering yield for different spike radii. MD
calculations appear as symbols whereas Hydrodynamics results are plotted as
thick lines. Thin, straight lines show the sputtering yield obtained using
Eq. (\protect{\ref{App3}}).}
\label{Y-for-R}
\end{figure}

In order to verify this simple expression, we calculate the sputtering yield
for different spike radii. The results, that appear in figure \ref{Y-for-R},
show that our fluid dynamics calculations compare fairly well with the MD
yields, and that Eq.(\ref{App3}) accounts reasonably well for the yields at
high-excitation energies. Discrepancies between MD and fluid dynamics at low
excitation energies and for small spike radii do appear. A detailed analysis
of such deviations was not carried out. As previously mentioned, the various
quantities entering our model do not accurately account for the
Lennard-Jones fluid in the MD simulations. In addition, having assumed the
solid target is a fluid, effects arising from the crystalline structure and
the atomic nature of the target can not be described. In the MD simulations
focused collision sequences play an important role at carrying energy and
momentum away from the spike, particularly for small spike radii. Finally,
it is worth noticing that equation (\ref{App3}) predicts a linear dependence
of the yield with the excitation or deposited energy. A result that was
derived in Ref.\cite{Kitazoe} using a simple, intuitive model rather than
well supported, rigorous calculation.

Although we have chosen not to address to the problem of crater formation,
it is worthwhile observing that late in our calculations craters do appear,
and they are all surrounded by a rim several $\sigma $ high (the reader is
referred to Ref.\cite{craters} for additional information about crater
formation). The hole left by the spike is normally greater than the initial
radius of the hot core. It is formed as the result of the net displacement
produced by the elastic wave along the radial direction. Near the edge of
the hole, the radial momentum is less than it is in the material below. As a
result, a kind of cantilever is formed which is pushed upwards by the fluid
below. See the case of $t$=2 ps in figure \ref{N-and-V-of-t(2)}. It is 
important to mention, however, that for $E_{exc}$ smaller than, say $U$, then
no hole is formed.

\section{Conclusions}
\label{Conclusions}

Sputtering at relatively high excitation densities is an old but unsolved
theoretical problem in ion solids interactions. Analytic diffusive thermal
spike models are commonly used to interpret data at high excitation
densities, although these models were never tested against more detailed
calculations. In addition, there is a history of applying ideas from fluid
dynamics to explain sputtering at high excitation density. These models are
called by a number of names (gas flow \cite{GasFlow}, shock \cite{Kitazoe},
pressure pulse \cite{PrssPulse} etc.) and attempt to account for the fact
that sputtering at high excitation density does not occur on an atom by atom
basis. These models also require a more detailed theoretical justification.

Establishing a theoretical basis for sputtering models at high excitation
density has been addressed recently by MD simulations using model materials
and simplified descriptions of the initial conditions. It was shown that
standard spike models break down at precisely those high excitation
densities which they were intended to treat. In addition, a new sputtering
regime was found. On increasing the energy density in the spike for fixed
spike radius, the yield changed from a non-linear dependence on the
excitation density to a linear dependence even though the ejection process
clearly remained non-linear. This is contrary to the conventional wisdom and
suggests saturation occurs in the sputtering. To examine this result we
first showed that such a regime is never attained for any set of material
properties using the diffusive thermal spike model \cite{MJakas}. Since the
standard spike model involves solving only the energy equation, we then
numerically integrated the full set of fluid equations for a 1D model of a
cylindrical spike \cite{PRB1}. Differences with the MD result remained which
we attributed to the lack of a surface. Here use a 2D fluid dynamics model
with a surface to confirm that when the full set of equations is treated the
MD result at high excitation density can be attained. Therefore, as pointed
out earlier, the principal deficiency of the standard spike model is the
assumption that the transport is diffusive.

We have calculated for the first time the sputtering yield from a
cylindrical thermal spike by directly integrating the full 2D fluid-dynamics
equations. The transport of mass and momentum are seen to play a significant
role in the ejection process. Since the conversion of the thermal energy
into kinetic/potential energy within the spike occurs very early, the
ejection process at high energy densities is much more closely related to an
``explosion'' rather than to the thermal diffusion and evaporation models 
\cite{S&C} typically used to describe sputtering at high energy densities.
Comparisons with MD-simulations using appropriate material parameters, show
that our fluid dynamics description can account for the main features of the
cylindrical thermal spike. These calculations also confirm the MD result
that transport along the cylindrical axis is as important as radial
transport and, therefore, a 2D model is required. We show the reported
nearly linear yield comes about because of the competition between
pressurized ejection and the transport of energy away from the spike by the
pressure pulse.

Using the evolution of the streamlines seen in these calculations we obtain
a simple expression for the yield at high excitation density for a
reasonable set of material parameters. Bringa and co-workers (\cite{BJJ}, 
\cite{BJJ2}) had shown that in this regime the yield could be written in the
form, $Y\approx C[R_{cyl}/l]^{m}\{[dE/dx]_{eff}(l/U_{0})\}^{p}$\ where 
$[dE/dx]_{eff}$ is the energy deposited that ends up fueling the spike (here 
$\pi R_{cyl}^{2}NE_{exc}$) and m and p are close to one. They gave $C\approx
0.18$\ for an LJ solid, which also appeared to apply to results for other
pair potentials \cite{BJJ}. Here we use a picture of the ejection attained
from the 2D fluid dynamics model to establish the theoretical basis for the
value of C. That is, the internal pressure in the spike determines a
critical radius ($R_{s}\approx R_{cyl}(Eexc/Uo)^{1/2}$) and a depth $\sim
R_{cyl}$, leading to the ejection of a conical volume of material $Y\approx 
\frac{1}{3}R_{cyl}\pi R_{s}^{2}N$. This gives $C=\frac{1}{3}$, which is
larger than the MD result. The difference is due in part to the fact that
the material properties are not exactly those of the LJ solid and transport
along crystal axes removes energy from the spike as discussed, however, all
the principal features of the transport and ejection are the same. This new
model resembles that of Yamamura and co-workers \cite{Kitazoe} but disagrees
with the `so-called' pressure pulse model used for molecular materials  
\cite{PrssPulse}.

Several points need further investigation. The formation of craters at
normal incidence is a topical problem that can be addressed by the model
developed here. Further, the connection between the sputtering yield and the
time used to heat the spike needs to be studied. In this paper, as in most
of the MD simulations, we assumed it to be negligibly small. This may be
correct for spike formation by a collision cascade, but is known to fail for
electronic sputtering of rare-gas solids \cite{johnson-schou}. Finally, it
must be noted that the fluid dynamics description of the spike is a useful
complement to MD. In the fluid model a broad range of material properties
and types can be readily studied, whereas complicated potentials are needed
in MD calculations of different materials. In fact it is seen in Figs \ref
{Y-for-lambda}-\ref{Y-for-eta} that the saturation leading to the linear
regime is not simply dependent on the cohesive energy ($E_{exc}\approx U_{0}$)
and the initial $R_{cyl}$, as found in the MD simulations using pair
potentials, but also depends on the material parameters $\lambda $ and $\eta
^{*}$. In addition, local equilibrium chemistry, which can play an important
role in many of the materials of interest to us, can be readily included
in the fluid models. However, MD has the advantage that non-normal incidence
can be treated easily, the state of the ejecta (clusters vs. atoms) can be
studied, and non-equilibrium chemistry can be introduced. Therefore, a
program in which complementary calculations using fluid dynamics and MD
simulations is underway. Here we have shown that a new linear sputtering
regime is seen in both models and we have developed a simple analytic model
for the yield at normal incidence.

\acknowledgments
Part of this work was carried out during a visit performed by one of the
authors (MMJ) to the School of Engineering and Applied Science, University
of Virginia. Financial aid from the Astronomy and Chemistry Divisions
of the National Science Foundation (U.S.A.) and the Consejer\'{i}a de
Eduaci\'{o}n, Cultura y Deportes del Gobierno Aut\'{o}nomo de Canarias
(Spain) are acknowledged.

\end{multicols}


\begin{references}
\bibitem[*]{byline}  Corresponding author. e-mail: {\it mmateo@ull.es}.

\bibitem{Spikes}  See for example C.T. Reimann. Mat.-fys. Medd {\bf 43},
Det. Kong. Dan. Vid. Selsk. 351 (1993) and references therein.

\bibitem{Thompson}  M. W. Thompson. Phil.Mag. {\bf 18}, 377 (1968).

\bibitem{Sigmund}  P. Sigmund. Phys.Rev. {\bf 184}, 383 (1969).

\bibitem{Bob-Evatt}  R.E. Johnson and R. Evatt. Radiation Effects, {\bf 52},
187 (1980).

\bibitem{S&C}  P. Sigmund and C. Claussen. J. Appl. Phys. {\bf 52}(2), 990
(1981).

\bibitem{andersen}  H.H. Andersen. Phy.Rev.Letters {\bf 80}, 5433 (1999).

\bibitem{johnson-schou}  R.E. Johnson and J. Schou Mat.-fys. Medd {\bf 43},
Det. Kong. Dan. Vid. Selsk. (1993).

\bibitem{Urbassek-Bob}  H.M. Urbassek, H. Kafemann and R.E. Johnson. Phys.
Rev. {\bf B 49}, 786 (1994-II).

\bibitem{Edu-Bob}  E.M. Bringa and R.E. Johnson. Nucl. Instr. Methods in
Phys. Res. {\bf B 152}, 167 (1999).

\bibitem{PRB1}  M.M. Jakas and E.M. Bringa. Phys. Rev. {\bf B 62}, 824
(2000).

\bibitem{MJakas}  M.M. Jakas. Radiation Effects and Defects in Solids, 
{\bf 152}, 157 (2000).

\bibitem{Landau}  L.D.D. Landau. Fluids Mechanics 2nd Ed.,
Butterworth-Heinemann (1995).

\bibitem{Zeldovich}  Yu. B. Zel'dovich and Yu. P. Raizer. Physics of shock
waves. Academic Press (1969).

\bibitem{HYDRO2D}  M.M. Jakas. {\em In preparation}.

\bibitem{BJJ}  E.M. Bringa, M.M. Jakas and R.E. Johnson. Nucl. Instr.
Methods in Phys. Res. {\bf B 164-165}, 762 (2000).

\bibitem{Hisrchfelder}  J.O. Hisrchfelder, C.F. Curtiss and R.B. Bird in 
{\em Molecular Theory of Gases and Liquids} John Wiley \& Sons, Inc. (1964).

\bibitem{c}  If the temperature of the spike is high, then, the speed of
sound ($c$) is greater than $c_{0}$ since $c=(p/MN)^{1/2}$, therefore
according to Eq.(\ref{Ptotal}),\ $c=\left( \lambda k_{B}T+c_{0}^{2}\right)
^{1/2}$.

\bibitem{U&S}  It must be noticed that, except for a numerical factor, Eq.(%
\ref{App2}) coincides with the {\em effective sputtering radius} derived by
H.Urbassek and P.Sigmund [Appl.Phys. {\bf A 33}, 19 (1984)] using a Gaussian
thermal spike. As they were obtained using different models such an
agreement appears to be a remarkable coincidence for which we have no
feasible explanation.

\bibitem{Kitazoe}  Y. Kitazoe, N. Hiraoka and Y. Yamamura. Surface Science 
{\bf 111}, 381-394 (1981).

\bibitem{craters}  Z. Insepov, R. Manory, J. Matsuo and I. Yamada. Phys.
Rev. {\bf B 61}, 8744 (2000).

\bibitem{GasFlow}  H.M.Urbassek and J. Michl. Nucl. Instr. and Methods in
Phys. Res. {\bf B22}, 480 (1987).

\bibitem{PrssPulse}  D.\ Feny\"{o} and R.E. Johnson. Phys. Rev. {\bf B 46},
5090 (1992-I)

\bibitem{BJJ2}  E.M. Bringa, R.E.Johnson and M.M. Jakas. Phys. Rev. {\bf B 60%
}, 15107 (1999)
\end{references}
\end{document}